\documentclass[aps,pra,preprint,superscriptaddress,showpacs,showkeys]{revtex4}

\usepackage{graphicx}
\usepackage{amssymb}

\begin{document}

\title{ Variations of the  Lifshitz-van der Waals  force  between metals immersed in liquids}

 \author{R. Esquivel-Sirvent}
\affiliation{Instituto
de F\'{i}sica, Universidad Nacional Aut\'{o}noma de M\'{e}xico \\
Ciudad Universitaria, D. F. 01000, M\'{e}xico.}
\email{raul@fisica.unam.mx}

 \date{\today}

\begin{abstract}
We present a theoretical calculation  of the Lifshitz-van der Waals  force between two metallic slabs 
embedded in a fluid, taking into account the change of the Drude parameters of the metals when in contact with  liquids of different  index of refraction.  For the  three liquids considered in this work, water,  $CCl_3F$  and $ CBr_3F$ the change in the Drude parameters of the metal imply a difference of up to $15\%$ in the determination of the force at short separations. These variations in the force is bigger for liquids with a higher index of refraction. 
\end{abstract}

\maketitle

\section{Introduction}
The Lifshitz theory prediction of  retarded van der Waals forces or Casimir forces has been tested with a high precision, beginning  in  the late nineties  \cite{lamoreaux,mohideen,decca,bressi} where the force was measured between a plate and a large sphere.    Several experiments have been made to test the dependence of the Casimir force with the dielectric properties of the materials. For example, semiconductors, systems whose dielectric function  changes with the environment or with external light sources, lateral Casimir forces, among others \cite{iannuzzi,lamoreaux2,mohideen2,mohideen3,chen08}.   
 
 In a new generation of experiments,  the Lifshitz-van der Waals  force was measured for two Au surfaces submerge in ethanol \cite{munday07}.   The measurements were done using an atomic force microscope with a fluid cell,  and constitute the first measurement of these forces in fluids.   Munday et al. later  presented an improved analysis of their data and a new set of experimental results taking into account electrostatic forces and Debye screening \cite{mundayrep,munday2}. In another experiment,  long range repulsive Casimir forces were measured between Au and silica immersed in bromobenzene \cite{munday3}, corroborating the prediction of the Lifshitz theory that the sign of the force can be change by a suitable combination of the dielectric functions of the materials. Experiments by   van Zwol $et$ $ al.$ also validated Lifshitz theory measuring the force between Au and glass immersed in alcohol using inverse colloid probe atomic force microscopy \cite{vanzwol}.
 
Even a subtle modification of the metallic surface can change the optical properties and thus the Lifshitz-van der Waals force, as has been shown by Ederth \cite{ederth} with Au surfaces modified with  self-assembled monolayers of alkylthiols.  Changes in the optical properties due to the metal-thiolate layer have been made using ellipsometric techniques even though the effective thickness of this layer is only a few Anstroms \cite{shi}.  

To compare with the experiments, Lifshitz \cite{lifshitz} theory for the Casimir force is used.    For  two slabs with a local dielectric function $\epsilon_1(\omega)$ and $\epsilon_2(\omega)$ separated by a gap of length $d$ filled with a dielectric function $\epsilon_3(\omega)$.
 Lifshitz result for the force per unit area is 
    \begin{equation}
       \label{lifshitz}
       F=\frac{\hbar c }{2 \pi^{2}}\sum_{\nu=s,p}\int_{0}^{\infty} d\zeta \int_{0}^{\infty}dQ Q k_3 \frac{r_{13}^{\nu} r_{23}^{\nu}}{e^{2k_3d}-r_{13}^{\nu} r_{23}^{\nu}},
       \end{equation}
       where $r_{ij}^{\nu}$ is the reflectivity between medium $i$ and $j$ for either $p$ or $s$ polarization, 
  $Q$ is the wave vector component along the
plates, $q=\zeta/c$ and $k_3=\sqrt{\epsilon_3 q^2+Q^2}$. The above expression is evaluated along the imaginary frequency axis by making the rotation to the imaginary axis of the frequency $ \omega\rightarrow i\zeta$.

The parameters needed in this expression are the separation between the plates and the dielectric functions $\epsilon_1, \epsilon_2, \epsilon_3$  to calculate the reflectivities.   The dielectric data in Casimir force measurements is commonly taken from tabulated data.  This has been proven to be a very important issue,  specially when comparing theory with experiment.  Detailed analysis of the optical properties performed by Pirozhenko \cite{irinaau} proved that  different tabulated data,  gave different values of the plasma frequency and damping parameter for Au.  This implied   variations in the Casimir force of up to 8\%.  Also, Svetovoy  et al.  \cite{george2} measured the optical properties of Au films of different thicknesses and preparation conditions showing again the importance of in-situ measurements. Depending on the sample, variations on the calculated Casimir force of up to 14\% were obtained.   For the experiments of the Casimir force in fluids,  similar issues appear. Large dispersion in the reported optical data of the fluids leads to a large dispersion in the calculated Casimir force, as 
discussed by Palasantzas \cite{georgedisp}.

In this paper we discuss another effect that changes the calculation of the Casimir force. This is,  the variation of the Drude parameters of metals when immersed in different fluids.  As we show, these changes in the Casimir force are within the resolution of current experiments and can be of help in getting a precise comparison between theory and experiment.

\section{Metals immersed in Liquids}

The experiments done in fluids assume that the optical properties of the metal do not change when in contact with the liquid.  
   However, as was shown experimentally by Gugger \cite{gugger} and by Chen \cite{lynch} when metals such as Ag and Au are immersed in liquids the measured Drude parameters of the metal change depending on the index of refraction of the fluid $n$.  Thus the use of the "dry" Drude parameters ($n=1$) can led to errors in the comparison between theory and experiment in Casimir force experiments.

The optical properties of metallic  films in the intra-band frequency region is described
 by a Drude model
\begin{equation}
\epsilon(\omega)=\epsilon_{\infty}-\frac{\omega_p^2}{\omega(\omega+i\gamma)}. 
\label{drude}
\end{equation}
  The damping parameter $\gamma$ is frequency dependent and is given by
\begin{equation}
\gamma=\gamma_0+\beta \omega^2,
\label{gamma}
\end{equation}
being $\gamma_0$ is the damping parameter at zero frequency and $\beta$ is a constant that depends on electron-phonon and impurity scattering \cite{smith82}.  For noble metals,  $\epsilon_{\infty}$ comes from the  core polarization due to shallow d-orbitals \cite{maier}.  

 Gugger et al. \cite{gugger} observed  that the measured Drude parameters for Ag films varied when the films were immersed in fluids.  Depending on the index of refraction $n$   different values of $\beta$, $\epsilon_{\infty}$, $\gamma_0$ and $\omega_p$ were obtained.  The method of total attenuated reflection was used in Ag films of approximately $500$ \AA  thick. These was independently confirmed by Chen and Lynch \cite{lynch} in Ag and Au films using spectroscopic ellipsometric measurements.  The films used by Chen and Lynch were 99.99\% pure Ag or Au films with a thickness varying from 1700 to 2700 \AA.  In this paper,  the data for the Drude parameters measured by  Chen and Lynch  are used. The parameters  are summarized in Table 1.

To use Eq.(2) we need to calculate the reflectivities and for this the dielectric functions $\epsilon(i\zeta)$ have to be determined.  The rotation to the complex axis is done using Kramer-Kronig relations
\begin{equation}
\frac{\epsilon(i\zeta)}{\epsilon_{\infty}}=1+\frac{2}{\pi \epsilon_{\infty}}\int_0^{\infty}d\omega \frac{\omega \epsilon"(\omega)}{\omega^2+\zeta^2},
\end{equation}
where $\epsilon"(\omega)$ is the imaginary part of the dielectric function given by Eq.(3), ( See for example Ref.\cite{Galuza01}).

The change in the dielectric function of  Au when it is in contact  with different fluids is shown in Figure (1).  We plot, according to Eq. (3),  $\epsilon(i\zeta)/\epsilon_D(i\zeta)$, being $\epsilon_D$ the dry dielectric function of Au and $\omega_{pD}$ its dielectric function to which all frequencies are normalized.  As the index of refraction increases, the changes on the Drude parameters of the  dielectric function of Au are more significant. 

To calculate the Casimir force the frequency dependent dielectric function of the liquids is needed. For water,  we use the  Parsegian-Ninham representation for the dielectric function that consists of a  sum of Lorentz-type dielectric functions 
\begin{equation}
\epsilon(i\zeta)=1+\frac{B}{1+\zeta \tau}+\sum_i \frac{C_i}{1+(\zeta/\omega_i)^2+g_i \zeta/\omega_i^2},
\label{ninham}
\end{equation}
where $B$,$C_i$ and $g_i$ are related to  the oscillator strength and $\omega_i$ is the resonance frequency.  The main contributions to the dielectric function of water come from the microwave, the infrared and the ultra violet region.  The spectral parameters needed to fit the dielectric function of water are obtained from Ref. (\cite{bergstrom}).

 The work of Lynch and Chen only specifies
that water ($n=1.33$), toluene ($n=1.51$), and several fluorocarbons were used in the experiments.    The index of refraction of the fluorocarbons used coincide with  $CCl_3F$ (n=1.42) and  $ CBr_3F$ (n=1.60). For these two liquids the dielectric function is determined from a single oscillator Cole-Cole model that after applying Eq.(5)  is
\begin{equation}
\epsilon(i\zeta)=\epsilon_0+\frac{\epsilon_{\infty}-\epsilon_0}{1+(\zeta \tau)^{1-\alpha}},
\end{equation}
where $\alpha$ is the distribution parameter, $\tau=1/\gamma$, and $\epsilon_0$ is the static dielectric function.  The parameters needed for $CCl_3F$  and $ CBr_3F$ are obtained from Ref. (\cite{buckley}).

The changes observed  in the dielectric function of metals in contact in liquids  is based on simple effective medium  approximations. The basic assumption is that the grain boundaries in the metal are infiltrated by the liquid. This makes that the surface acts as a composite medium of liquid and metal, with an average or effective dielectric function $<\epsilon(\omega)>$.  This effective function can be calculated using Bruggeman's effective medium theory \cite{bruggeman}
  \begin{equation}
  f_M \frac{\epsilon_M-<\epsilon>}{\epsilon_M+2<\epsilon>}+f_F \frac{\epsilon_F-<\epsilon>}{\epsilon_F+2 <\epsilon>}=0,
  \label{effective}
  \end{equation}
where $f_{M,F}$ is the volume fraction of metal or liquid and $\epsilon_{M,F}$ their dielectric function.  This same system was used by Palasantzas $et$ $ al.$ \cite{vanzwol2} to explain the Casimir force between metal surfaces when an ultra thin film of water was deposited on their surface.   

\section{Casimir force calculation}
The main purpose of this paper is to study the effect on the Casimir force calculations due to the variations of the Drude parameters of the metal when in contact with different liquids. To calculate the Casimir force, we consider two Au slabs separated a distance $L$ and the gap between the slabs is filled with a liquid that can be water , $CCl_3F$   or $ CBr_3F$. 
Using Eq.(2) we compute the force $F_D$ between the two Au slabs 
assuming that the Drude parameters of Au do not change, and are the same as in the dry sample and compare it to the force $F_m$ obtained when we use the modified parameters 
of Au  shown in Table 1. The percent difference 
\begin{equation}
\Delta \%= |\frac{F_D-F_m}{F_D}|\times 100,
\end{equation}
 will show the effect due to  the variations of the Drude parameters of Au in contact with  different liquids. 
 In Figure 2 we plot the percent difference as a function of slab separation for water and two of the fluorocarbons.  At short separations the percent difference between the  Lifshitz theory using the unmodified Drude parameters and the modified ones is larger. At large separations the difference becomes smaller, and eventually converges to the same value, however the percent difference $\Delta\%$ never go to zero. The higher difference at short separations is for  $ CBr_3F$ that has the higher index of refraction in the visible. For this fluid, the apparent plasma frequency of Au is larger making it a better reflector and thus increasing the Casimir force.  
  
  \section{Conclusions}
  
  In this paper we show that the measurements of the Drude parameters of metals in different fluids affect the calculation of the Casimir force and the variations of the force between "dry" and "wet" Au can led to difference of up to $15 \%$.   
 Although we can not compare directly with the experiments of Munday \cite{munday07, munday2,munday3} since the Drude parameters for Au will have to be measured in ethanol or bromobenzene,   the results of this paper  show that the effect on the Casimir force due to the variation of the Drude parameters can be detected experimentally within the current experimental techniques.  The variation of the Drude parameters in different fluids show that the accurate determination of the dielectric function is important,  in particular given the broad applications now envisioned for the Casimir effect, such as Casimir chemistry
\cite{Sheehan09}.
\acknowledgments {Partial support from DGAPA-UNAM project IN 113 208.}

\newpage

\begin{table}[h]
	\begin{tabular}{|c|c|c|c|c|}
	\hline
		n    & $\epsilon_{\infty}$ & $\omega_p^2 (eV^2)$ & $\gamma_0(eV)$&$\beta (eV^{-1})$  \\
		\hline 
		 1& 7.76 & 71.53 & 0.0041 & 0.0123 \\
		 1.33& 8.71 &  79.97& 0.0049 & 0.0153 \\
		1.42 & 9.17 & 82.52&0.0062  & 0.0055 \\
		 1.51&9.65  &  85.60& 0.0066 &0.0059  \\
		 1.60& 10.30 &88.33  & 0.0097 &  0.0072 \\
		 \hline
	\end{tabular}
	\caption{Modification of the Drude parameters of Au for different liquids of index of refraction $n$. Data taken from Ref.(\cite{lynch}).}
	\label{table1}
\end{table}

\newpage

 \begin{figure}
 \includegraphics[width=12cm]{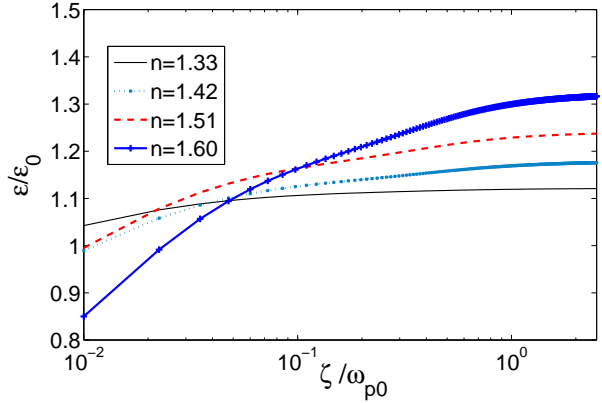}
 \caption{Dielectric function of Au slabs in contact with different liquids. The dielectric function has been rotated $\omega\rightarrow i\zeta $ and is normalized to the dielectric function of Au in air $\epsilon_D$ ($n=1$ in Table 1). The   corresponding plasma frequency is $\omega_{pD}$. (color online)}
 \end{figure}

 \begin{figure}
 \includegraphics[width=8cm]{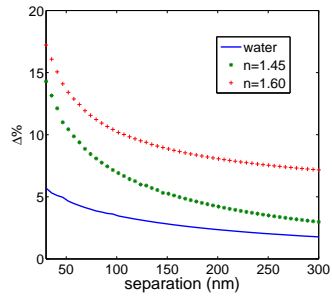}
 \caption{Effect of the variations of the Casimir force for three different liquids: water ($n=1.33$),  $CCl_3F$( $n=1.42$)  and $ CBr_3F$ ($n=1.66$). The index of refraction are used as a guide, the calculation of the force took into account the frequency dispersion of the liquids according to Eq. (6) and Eq. (7).  The vertical axis shows the percent difference $\Delta \%$  between the force for the modified and unmodified  Drude parameters of  Au.  The shortest distance between the plates in the graph os $10$ nm. (color online)} \end{figure}

\end{document}